\documentclass[iop]{emulateapj}
\usepackage{hyperref}
\usepackage{color}
\usepackage[utf8]{inputenc}
\usepackage{natbib}
\usepackage{graphicx}
\usepackage{xfrac}
\usepackage{ulem}

\begin{document}
\title{
On the limitations of statistical absorption studies\\ with the Sloan Digital Sky Surveys I--III}
\author{Ting-Wen Lan\altaffilmark{1}, 
Brice M\'enard\altaffilmark{1,2},
Dalya Baron\altaffilmark{3}, 
Sean Johnson\altaffilmark{4,5,8}, \\ 
Dovi Poznanski\altaffilmark{3},
J.~Xavier Prochaska\altaffilmark{6}, and
John M. O'Meara\altaffilmark{7}
}
\shorttitle{Limitations of absorption studies with the SDSS}
\shortauthors{Lan et al.}

\altaffiltext{1}{Kavli IPMU, the University of Tokyo (WPI), Kashiwa 277-8583, Japan}
\altaffiltext{2}{Department of Physics \& Astronomy, Johns Hopkins University, 3400 N. Charles Street, Baltimore, MD 21218, USA}
\altaffiltext{3}{School of Physics and Astronomy, Tel-Aviv University, Tel Aviv 69978, Israel}
\altaffiltext{4}{Department of Astrophysical Sciences, 4 Ivy Lane, Princeton University, Princeton, NJ 08544, USA}
\altaffiltext{5}{The Observatories of the Carnegie Institution for Science, 813 Santa Barbara Street, Pasadena, CA 91101, USA}
\altaffiltext{6}{ Department of Astronomy and Astrophysics, UCO/Lick Observatory, University of California, 1156 High Street, Santa Cruz, CA 95064, USA}
\altaffiltext{7}{Department of Chemistry and Physics, Saint Michael's College, One Winooski Park, Colchester, VT 05439, USA 0000-0002-7893-1054}
\altaffiltext{8}{Hubble \& Carnegie-Princeton Fellow}

\begin{abstract}
We investigate the limitations of statistical absorption measurements with the SDSS optical spectroscopic surveys. We show that changes in the data reduction strategy throughout different data releases have led to a better accuracy at long wavelengths, in particular for sky line subtraction, but a degradation at short wavelengths with the emergence of systematic spectral features with an amplitude of about one percent. We show that these features originate from inaccuracy in the fitting of modeled $F$-star spectra used for flux calibration. The best-fit models for those stars are found to systematically over-estimate the strength of metal lines and under-estimate that of Lithium. We also identify the existence of artifacts due to masking and interpolation procedures at the wavelengths of the hydrogen Balmer series leading to the existence of artificial Balmer $\alpha$ absorption in all SDSS optical spectra. All these effects occur in the rest-frame of the standard stars and therefore present Galactic longitude variations due to the rotation of the Galaxy. We demonstrate that the detection of certain weak absorption lines reported in the literature are solely due to calibration effects. Finally, we discuss new strategies to mitigate these issues.

\end{abstract}


\section{Introduction}

Large spectroscopic surveys provide samples of spectra offering appealing opportunities for statistical analyses. They enable the detection of signals buried in measurement noise or population variance which, in some cases, are unreachable by targeted observations on large telescopes. When performing such analyses or when evaluating the statistical potential of spectroscopic surveys, it is important to understand the limits inherent to the data. In the optical, the Sloan Digital Sky Surveys I, II, and III \citep[SDSS,][]{York2000, Eisenstein2011} have been an important source of spectroscopic data for statistical analyses for almost two decades. Considering only the collecting power, combining a hundred SDSS spectra is comparable to a one night spectroscopic exposure of one object on an 8-meter telescope. The availability of millions of spectra allows one to detect weak spectroscopic features distributed over a large area or a wide redshift range -- as long as systematic effects are controlled. Statistical absorption measurements with SDSS data have led to the detection of numerous signals of astrophysical interest: the cosmic baryon acoustic oscillation feature \citep[e.g.,][]{Busca2013,Bautista2017}, weak absorption by hydrogen and metals (measured over three orders of magnitude in equivalent width) in the circumgalactic medium \citep[e.g.,][]{Nestor2003,York2006, Pieri2014,ZhuCa2013, Zhu2014, Sardane2015, Lan2017, Zhang16, Mas2017}, gaseous metals in the ISM \citep[e.g.,][]{Dovi2012, Brandt2012, Murga2015} and the characterization of the weak diffuse interstellar bands \citep[e.g.,][]{Lan2015,Baron2015,Baron2015b}. 

Extensive efforts have been made to optimally extract 1D spectra from the 2D spectra \citep[e.g.,][]{Smee2013, Bolton2010}, calibrate them \citep[e.g.,][]{Schachter1989, Schachter1990, Bessell1999}, and improve the subtraction of the correlated sky emission lines \citep[e.g.,][]{Wild2005}. In the context of statistical spectroscopic studies of the SDSS I-II datasets, limitations due to calibration and systematic effects have already been reached and/or reported by a number of authors \citep{Yan2011, Zhu2013, Lan2015}. In this study, we investigate absorption systematics in the context of the SDSS I, II, and III optical surveys, i.e. the main sample from SDSS I-II and the BOSS survey from SDSS III \citep{Dawson2013}. We show how these systematic effects have evolved through different data releases. Our results are informative for the design and optimization of ongoing and future spectroscopic surveys such as  SDSS-IV and V \citep{Blanton2017, Kollmeier}, DESI \citep{Schlegel2011,Levi2013}, PFS \citep{Takada2014} as well as surveys using integral field units \citep[e.g. MaNGA,][]{Bundy2015}. The SDSS targets F-type subdwarfs as standard stars. For simplicity, we refer them as F stars in the paper.

\section{Data analysis}

\subsection{Absorption spectra}
Spectroscopic surveys provide us with a set of flux density measurements as a function of wavelength, angular position and time: $\rm{F_{\rm meas}}(\lambda, \phi, t)$. For the SDSS survey, the flux density is sampled over a constant $\delta\lambda/\lambda$ with an angular aperture of a few arcsecond across. We are usually not interested in raw measured flux $\rm{F_{\rm meas}}(\lambda)$ but in calibrated flux ${\rm F_{\rm source}}(\lambda)$. A simple description of the dominant effects can be captured by the relation
\begin{equation}
    F_{\rm meas}= ( F_{\rm source} + F_{\rm sky} ) \times R \;,
\label{eq:R}
\end{equation}
where $F_{\rm sky}$ represents the additive flux contribution from the atmosphere and $R$ is the response function of the observing system, which includes the detector and telescope efficiency and the atmospheric transmission, quantities which depend on $\lambda$, $\phi$ and $t$. As the main goal of our study is to investigate the accuracy of absorption measurements, we will for simplicity assume that $F_{\rm sky}$ can be subtracted adequately and there are no other additive sources of light to be present. 
We will also assume that it is possible to estimate the large-scale continuum spectral density, $\hat F_c$, of an ensemble of sources on the sky, where $i$ denotes a given object:
\begin{equation}
\label{eq:delta}
\delta_F^{(i)}(\lambda,\phi) = \frac{F^{(i)}(\lambda,\phi)} {\hat F_c^{(i)}(\lambda)}\;.
\end{equation}
If $\hat F_c$ has been estimated using only large-scale information, then $\delta_F$ carries the imprint of the small-scale properties of the response of the system, $R(\lambda)$. We now have an ensemble of flux residuals or an ensemble of vectors
\begin{equation}
\left\{\delta_F^{(i)}(\lambda,\phi)\right\}_i\;.
\end{equation}
Investigating their statistical properties can reveal the presence of systematic effects. Beyond an inspection of the mean absorption spectrum, one can explore the origin of potential spectral features using dimensionality reduction techniques to gain insight into the nature of certain features as well as reveal the existence of effects undetectable in the mean but contributing to the overall higher-order moments. With SDSS-type spectra, the variance is typically dominated by  measurement noise. As a result, a principal component analysis does not reveal any insightful eigen vector. Variations around the mean can be characterized by cross-correlations with observables or spectroscopic templates.

\subsection{The mean SDSS absorption spectra}

We now investigate the limitations of statistical absorption measurements with different data releases of the SDSS surveys. To first define flux fluctuation spectra $\delta_F^i(\lambda)$ (Eq.~\ref{eq:delta}), we need to use sources for which we can robustly estimate the continuum flux. To do so we make use of SDSS-I/II quasar spectra from data release 7 \citep[DR7,][]{Abazajian2009, Schneider2010} and SDSS-III quasar spectra from data release 9 \citep[DR9,][]{Ahn2012,Paris2012} and 12 \citep[DR12,][]{Alam2015,Paris2017} of the Baryon Oscillation Spectroscopic Survey \citep[BOSS,][]{Dawson2013}. 
\citet{Zhu2013} estimated the continuum flux of virtually all the quasars observed by these surveys using a data-driven technique. They used non-negative matrix factorization (NMF), a dimensionality reduction technique, and a running median filter with 71 pixels in size to model the rest-frame spectral energy distributions (SEDs) of the entire population of quasars. In our analysis we use the sample provided by  \citet{Zhu2013}\footnote{The catalogs can be found at \url{http://www.guangtunbenzhu.com/jhu-sdss-metal-absorber-catalog}.}. 
After decomposing quasar spectra onto the NMF basis set, the authors selected quasars with eigen-coefficients within 5$\sigma$ of the mean eigen-coefficients of all the input quasars. In addition, we restrict the sample to quasars with a redshift lower than 2.2 to avoid the large variance from the Lyman alpha forest. This selection yields about 74,000 DR7 quasar spectra, 6,100 DR9 quasar spectra and 25,000 DR12 quasar spectra. We note that although the DR9 quasars are all included in the DR12 sample, the two samples of quasar spectra are reduced by two different versions of the pipeline.

\begin{figure*}
\center
\includegraphics[width=0.98\textwidth]{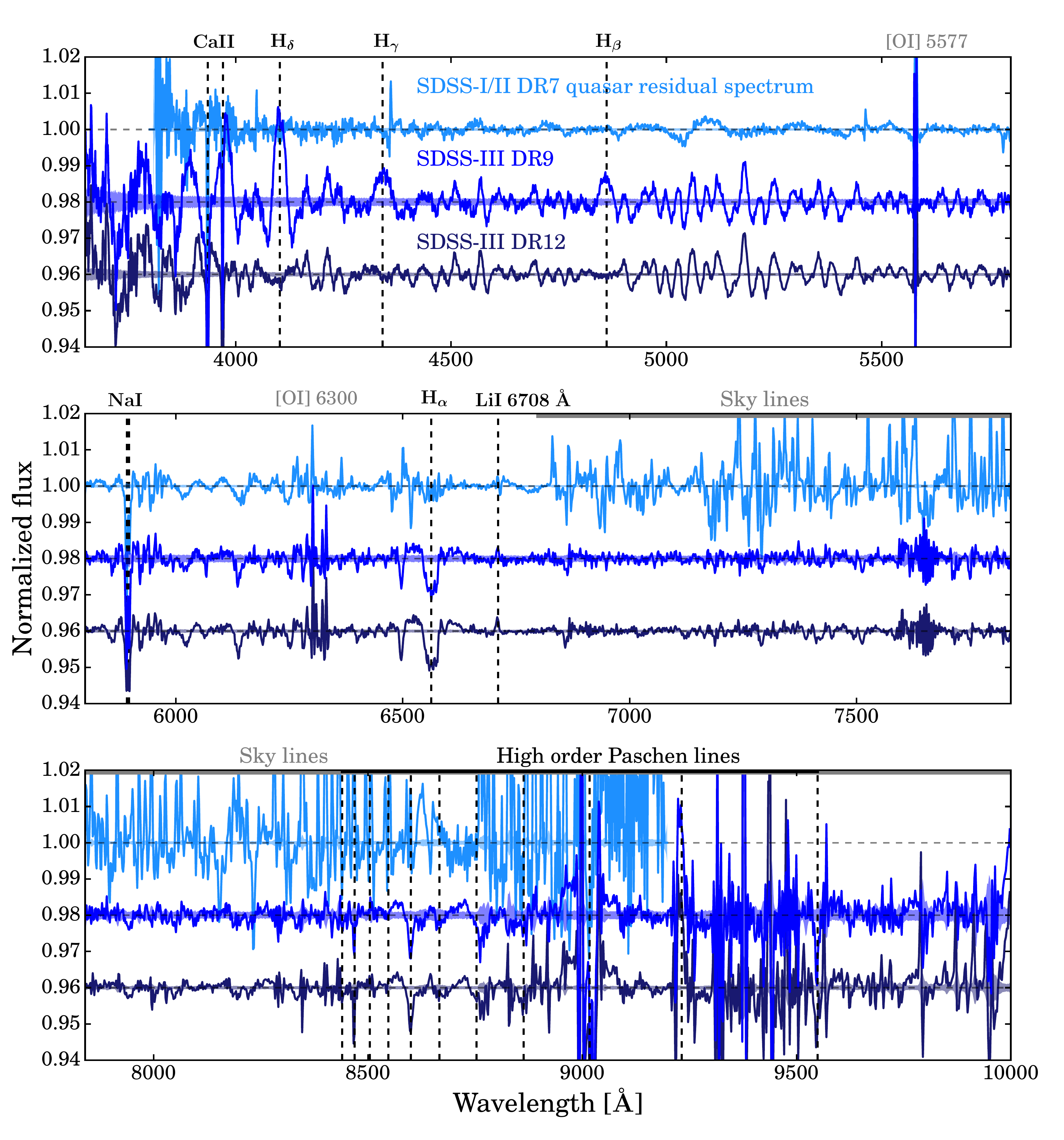}
\caption{Median residual spectra from SDSS I-II DR7, SDSS-III DR9 and SDSS-III DR12 data releases of quasar spectra. Several wiggles around 4500-5800 $\rm \AA$ and 6500 $\rm \AA$ appear in DR9 and DR12 composite spectrum but not in DR7 spectrum. The $1 \sigma$ uncertainty of each pixel is shown with shaded bands.}

\label{fig:composite_spectrum}
\end{figure*}

\begin{figure*}[t]
\includegraphics[width=\textwidth]{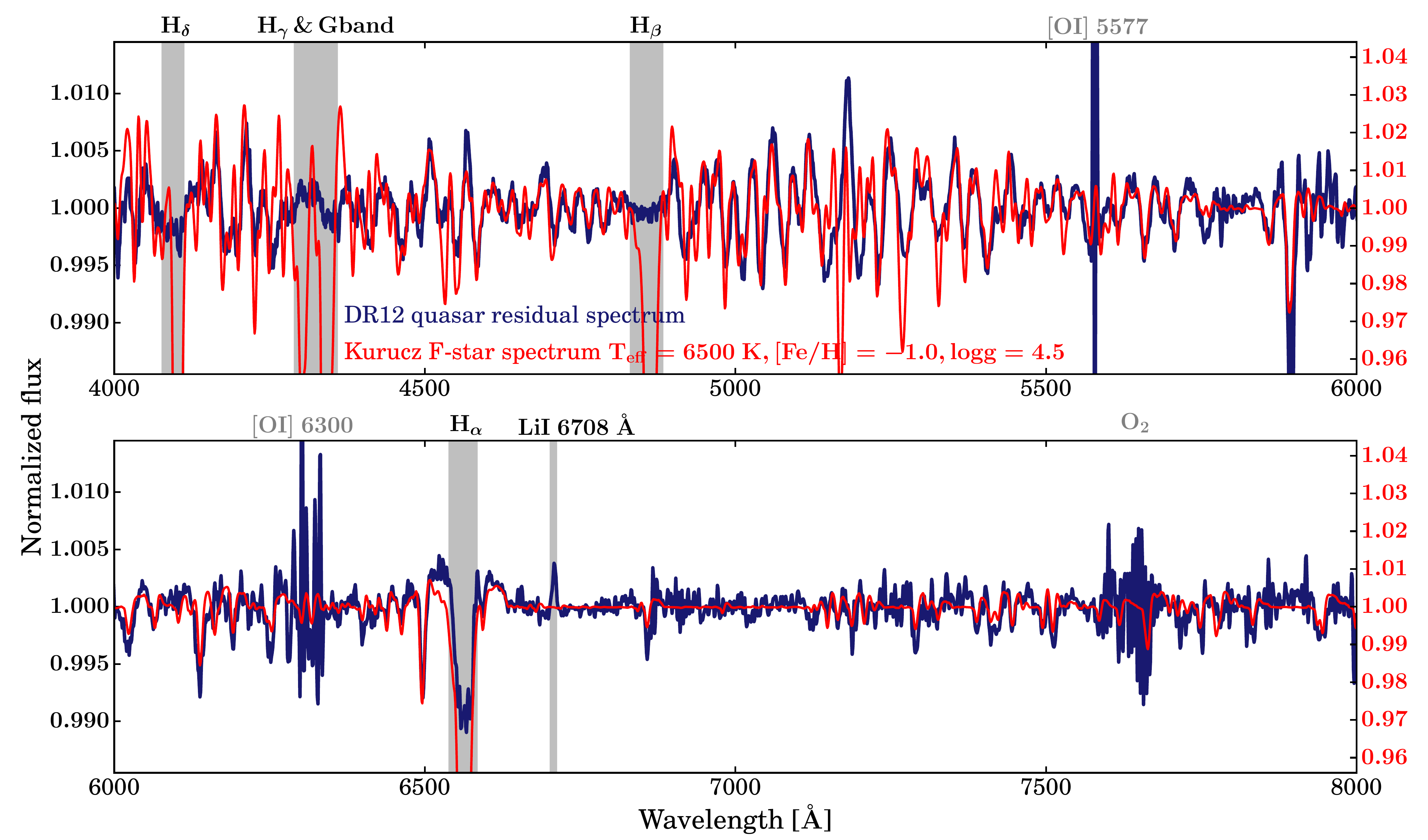}
\caption{Comparison between DR12 median quasar residual spectrum in blue and F-star model spectrum in red. The F-star model spectrum is in the scale shown on the right y-axis. The scale is chosen to illustrate the consistency of features in two spectra.}
\label{fig:comparison}
\end{figure*}

We first compute the mean and median absorption spectra $\langle \delta_F \rangle(\lambda)$ for each dataset in the observer frame to look for the presence of low-level systematics. We estimate the corresponding uncertainty of each pixel by bootstrapping each quasar sample 200 times.
The two estimators lead to consistent results. As expected, the median presents a lower noise level. We will therefore focus on it in the remainder of the paper.
The median residual spectra are shown in Figure 1. Ideally they should be consistent with unity everywhere but a number of significant departures can be seen. In particular, we observe:
\begin{itemize}
    \item The presence of the Calcium H\&K absorption lines  at $\lambda\lambda 3934, 3969$ in both datasets. Residuals due to Sodium D $\lambda\lambda 5891, 5897$ absorption lines are present. They are mainly due to absorption by the ISM of the Milky Way. This effect is expected as we are using extra-galactic objects as background sources. First discovered by \citet{Hartmann1904} and \citet{Heger1919}, these absorption lines have recently been  studied by \citet{Dovi2012} and \citet{Murga2015} and used to map out the distribution of interstellar and circumgalactic matter across the sky.
    \item At intermediate wavelengths, $4500<\lambda<7000\,$\AA, a number of wiggles appear with significance greater than $10 \sigma$. They are more numerous in the DR9 and DR12 composite spectra than in the DR7 composite spectrum, indicating that the quality of the data reduction has decreased with time in this region of the spectrum. The amplitude of the wiggles reaches up to $1\%$ of the continuum level. They are not related to the quasar SEDs as those objects span a wide redshift range. These systematic features most likely originate from the calibration process of the data.
    \item An absorption-like feature at the wavelength of $H{\alpha}$ is visible in DR9 and DR12 but not in DR7. We note that \citet{ZhangHa} \citep[see also][]{Sethi2017} interpreted the presence of this DR12 spectroscopic feature as $H{\alpha}$ absorption from the Milky Way halo. We will demonstrate below that this feature is a calibration artifact. It is also worth pointing out that we do not observe any absorption feature at the wavelength of $H{\beta}$. 
    \item An emission-like feature is seen at the wavelength of Lithium $6708\,$\AA.
    \item Broad emission-like features are seen at $H{\beta}$, $H{\gamma}$, and $H{\delta}$ in DR9 but not DR12. 
    \item At $\lambda>7000\,$\AA, we observe residuals due to OH sky line subtraction. They are substantially weaker in DR9 and DR12 compared to DR7, showing an improvement of the sky model of the data reduction pipeline. 
\end{itemize}
In short, from DR7 to DR12, the quality of the data reduction has improved in the red but degraded in the blue.

We now focus on the bluer part of the spectrum, devoid of OH sky emission lines and where the amplitude of the wiggles has increased from DR7 to DR12. An informed inspection of the wiggles seen for the DR12 data at intermediate wavelengths, $4500<\lambda<7000\,$\AA, shows that many of these fluctuations coincide with the position of absorption lines in F-star spectra. This is shown in Figure~\ref{fig:comparison} where we superpose the DR12 median absorption spectrum and the small-scale structure of a F-star model spectrum from \citet{Kurucz1992} obtained by removing a running median filtered estimate of the flux. Finding such a signature is expected at some level as SDSS targets such stars and uses them as spectral calibration standards. However we also observe that some features do not show the same level of similarity with the model F-star spectrum. These differences are highlighted with shaded regions in the figure. In order to understand the origin of all these features, we need to dive deeper into the calibration processes of the SDSS surveys.

\begin{figure*}
\includegraphics[width=1.02\textwidth]{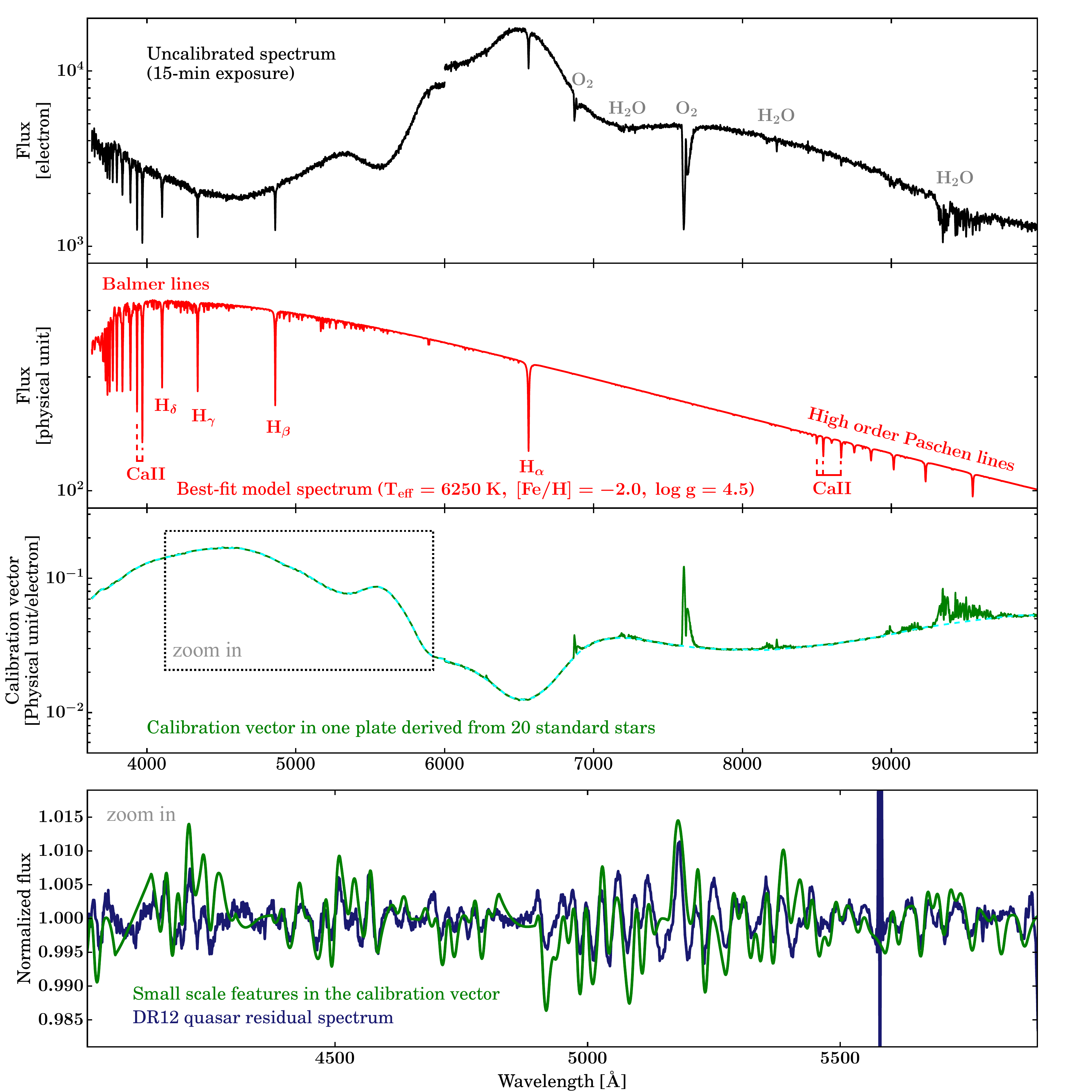}
\caption{Demonstration of SDSS calibration process. Panel one: an example of un-calibrated spectrum. Panel two: the best-fit model spectrum. Panel three: the derived calibration vector in the plate 4342. Panel four: small scale residuals in the calibration vector.}
\end{figure*}

\subsection{The SDSS spectroscopic calibration process}

\begin{figure*}
\includegraphics[width=\textwidth]{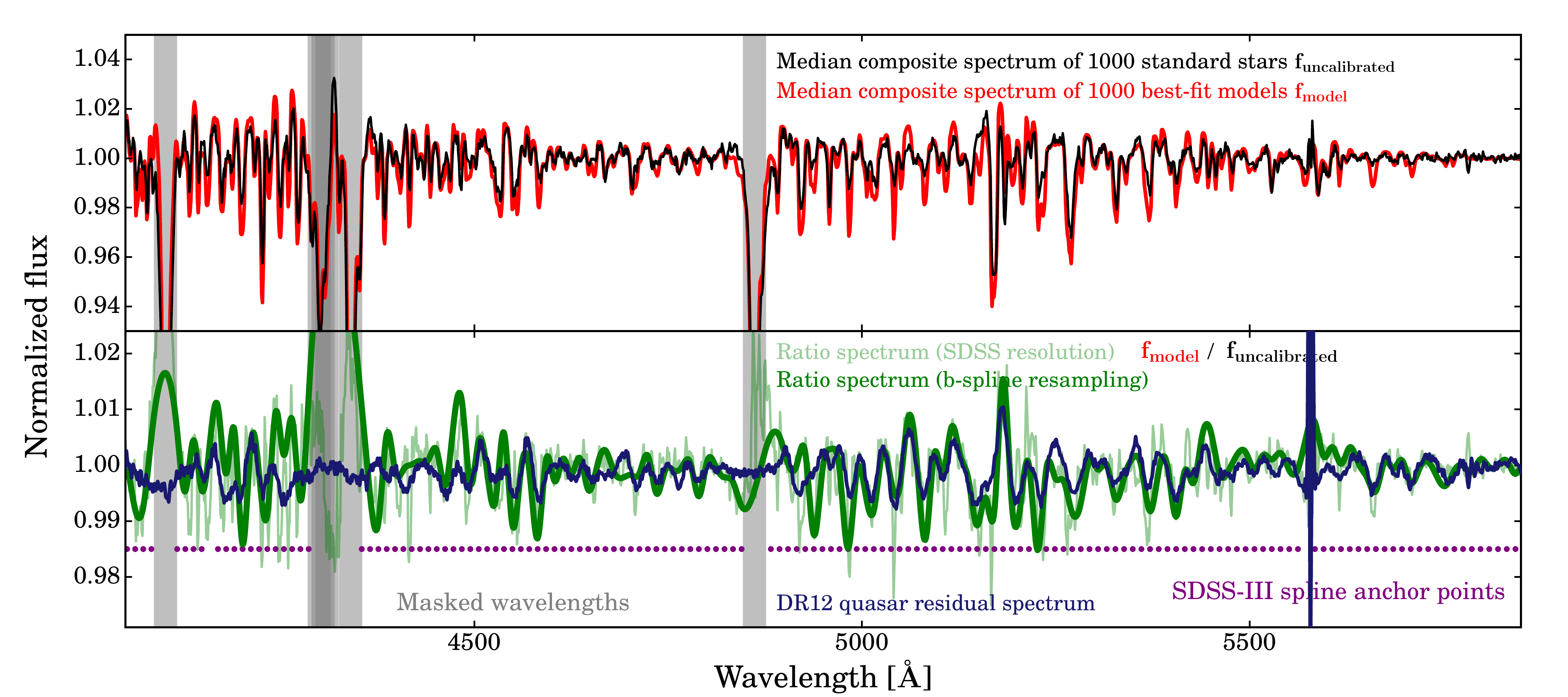}
\caption{Difference between observed stellar absorption lines and best-fit model spectra. Upper panel: median composite spectrum of 1000 standard stars derived from uncalibrated spectra (black) and composite spectrum of the corresponding best-fit model spectra (red). Lower panel: Ratio between the model and un-calibrated spectra with SDSS full resolution  (thin green line) and {a smooth version with a cubic b-spline re-sampling (thick green line) using SDSS-III anchor points (purple points)} and DR12 median residual spectrum (blue). The difference between the uncalibrated and model spectra is consistent with the residuals observed in quasar spectra.}
\label{fig:ratio}
\end{figure*}

We now review the SDSS spectroscopic calibration process. The relevant steps are described in \citet{Yan2016} (their Section 2.2). We focus on the DR12 release as it is the data reduction displaying the stronger systematic fluctuations. Here we summarize the  relevant key steps involved in the flux calibration procedure after instrumental effects, i.e. flat field and bias, have been corrected:
\begin{enumerate}
\item For each spectroscopic plate, SDSS-I / SDSS-III observes 16/20 standard stars which are selected to have colors consistent with the SDSS primary standard BD+17 4708 (F8 subdwarf)\footnote{The color selection can be found at \url{http://www.sdss.org/stdstar_colors/}} \citep{Abazajian2004,Dawson2013, Yan2016}.
For each flux-uncalibrated standard star spectrum, a running median filter with 99 pixels in width is applied to remove large scale fluctuations, mainly due to the response function $R(\lambda)$ in Eq.~\ref{eq:R}. A normalized spectrum with small-scale information is obtained and the apparent velocity of each standard star is estimated.
\item The normalized, small-scale spectrum is then fitted by a series of normalized theoretical spectra of F stars from \citet{Kurucz1992} with ranges of effective temperature, surface gravity, and metallicity\footnote{The data of the theoretical spectra can be found at \url{https://svn.sdss.org/public/repo/eboss/idlspec2d/tags/v5_7_0/templates/kurucz_stds_raw_v5.fits}}. This is done {\it at the velocity of the standard star}. A best-fit theoretical spectrum with minimum reduced $\chi^{2}$ is then selected. 
\item A calibration vector is derived by taking the ratio between the flux of the best-fit model and the uncalibrated flux of the standard star at the velocity of the observed standard star:
\begin{equation}
    \rm calibration \ vector = {R(\lambda)^{-1}} = \frac{F_{\rm model}(\lambda)}{F_{\rm meas}(\lambda)}.
\end{equation}
The pipeline then combines individual calibration vectors with S/N weighting and samples them using an iterative, multi-scale b-spline interpolation scheme to derive the final calibration vector. This effective scale should be chosen 
carefully to efficiently capture the response function of the instrument and avoid over-fitting small scale residuals due to the difference between modelled spectra and observed ones. We will discuss the effect of this b-spline interpolation scale on the final calibration vectors below. All the spectra in the same plate are then normalized by the final calibration vector. 
\end{enumerate}
The code performing this calibration procedure is available online\footnote{The main code for this procedure can be found at \url{https://svn.sdss.org/public/repo/eboss/idlspec2d/tags/v5_7_0/pro/spec2d/spflux_v5.pro}}.
Since the calibration vector is derived from the ratio between the uncalibrated and theoretical spectra, any mismatch between the two will generate features which will be imprinted onto all the science spectra generated by the survey. In other words, it is important to characterize the level of bias present in the calibration vector. 

To provide some visual guidance with the calibration process, we show the different steps involved in Figure~3 using a representative standard star. The large-scale features seen in the top panel are due to modulations in the sensitivity of the system as well as absorption bands originating from the atmosphere. The discontinuity visible at $\lambda\sim6000\,$\AA\ indicates the transition between the blue and red detectors of the spectrograph. The second panel of the figure shows the corresponding best-fit model spectrum (red). The calibration vector derived by taking the ratio of the two is shown in the third panel with the green line. Since we are interested in small scale features, we filter out the large-scale features using a cubic b-spline with break points separated by $50\,$\AA\ and avoiding the atmospheric absorption bands. This continuum estimate is shown by the cyan dashed line. The fluctuations around that baseline are shown in the bottom panel of the figure where we focus on the bluer part of the wavelength range, with $\lambda<6000\,$\AA. They reveal limitations in the ability of the best-fit model to reproduce the amplitude of the stellar absorption lines. For this one plate, the observed residuals appear to be similar to those visible in the median DR12 absorption spectrum. This shows that, already at the level of a single plate, detectable systematics are introduced by the calibration process. 

To illustrate the origin of those systematic features, we consider an ensemble of 1000 bright reference stars, randomly selected from the DR12 sample. For each object, we remove the large scale features due to the response function of the spectrograph with a running median filter and stack the normalized uncalibrated spectra with a median estimator. We apply the same procedure to the best-fit model spectra of these reference stars. The upper panel of Figure~4 shows the composite spectra of the standard stars in black and their corresponding best-fit models in red. The bottom panel shows the ratio with full SDSS spectral resolution (thin green line). This figure shows that, on average, {\it{the best-fit models tend to have stronger absorption features than those of the observed standard stars}}. This could be due to an inability to converge to the right metallicity, the use of an inappropriate point spread function, or/and the inaccuracy of the stellar models and atomic data for modelling the absorption lines \citep[e.g.][]{Kurucz1993,Heiter2015}. It is reasonable to think that this step of the calibration process could be further improved. This currently limits the accuracy of statistical studies with SDSS spectra at the plate level.

We now look more closely at the amplitude of the fluctuations. Figure~\ref{fig:ratio} shows that the features in the SDSS resolution ratio spectrum (thin green line) appear to be narrower and have a larger amplitude. This difference is even more severe if we consider the DR7 composite absorption spectrum. This is due to the fact that the SDSS pipeline does not use the calibration vector at full resolution but subsamples it and then uses a spline interpolation to define it at every pixel. We perform a similar analysis as the pipeline by using the anchor points of SDSS-III (purple data points) to resample the full resolution ratio spectrum and we find a good agreement between the smooth ratio spectrum (thick green line) and the DR12 composite absorption spectrum. We have also found that the number of anchor points for the spline interpolation is lower for DR7 than DR12. As a result, the effective calibration vector used in DR7 is intrinsically smoothed and this explains why the residual features in the DR7 composite absorption spectrum shown in Figure~1 are much weaker. 

Another limitation of the best-fit model is seen by the presence of an emission-line feature at the location of LiI 6708 $\rm \AA$ \citep[e.g.][]{Gerbaldi1995,Takeda2005} in the composite residuals. This shows that the best-fit models under-predict the strength of this line.
This could be due to an adoption of an inaccurate Li abundance or/and an adoption of incomplete stellar physics in the model \citep[e.g.,][]{Kurucz1996}.
As a result, LiI is under-estimated in all reduced spectra. All these systematic features, whether they are under- or over-represented in the best fit F-star model spectra,  {\it will imprint a systematic pattern in all the spectra of the dataset.} This pattern also reflects the velocity distribution of the standard stars across the sky and therefore carries their velocity distribution.

\begin{figure}
\includegraphics[width=0.5\textwidth]{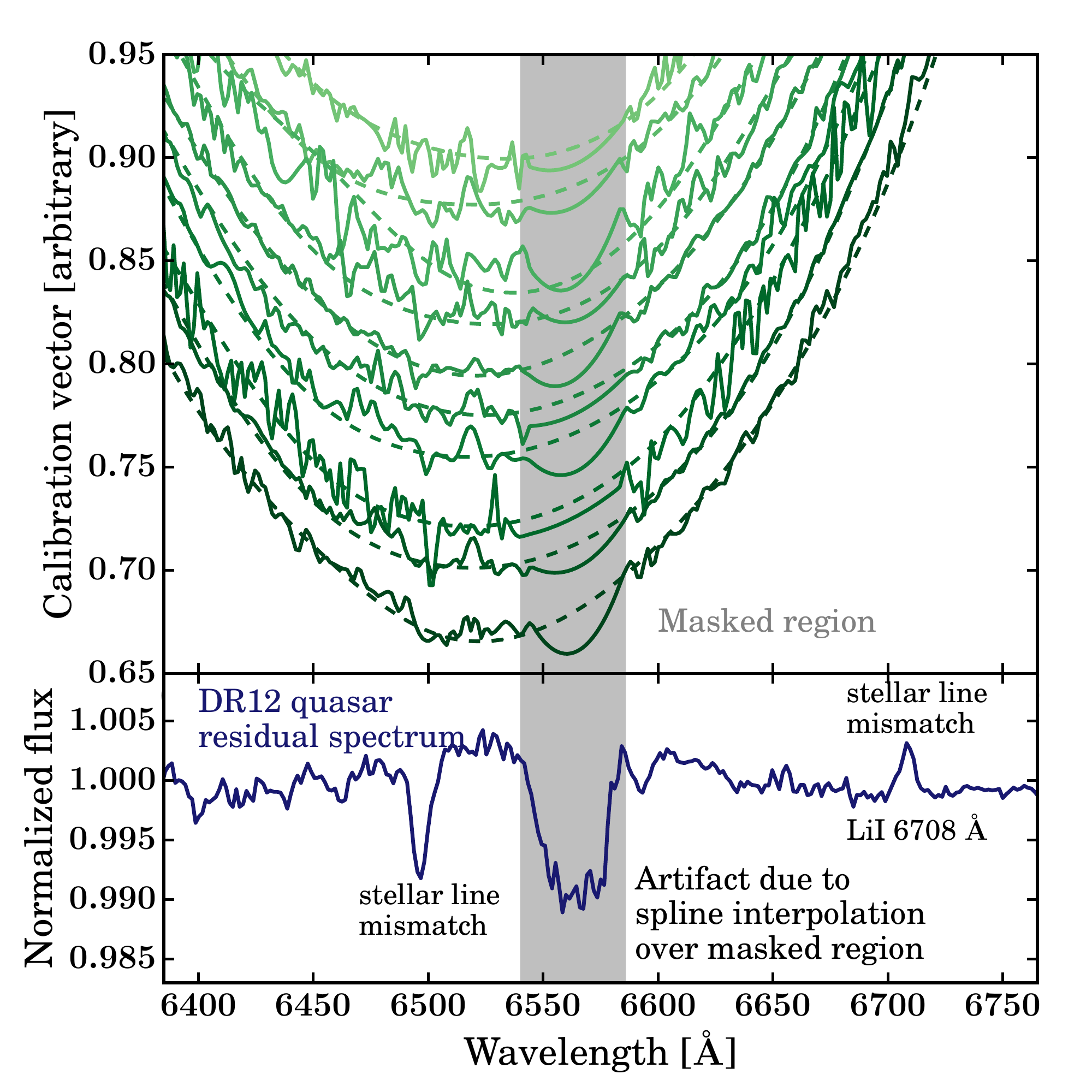}
\caption{Origin of the broad $H{\alpha} $ absorption. Upper panel: 10 randomly-selected DR12 calibration vectors around 6500 $\rm \AA$. Dashed lines show the large-scale features modeled by cubic b-spline functions
with break points separated by 50 $\rm \AA$. Lower panel: the DR12 quasar residual spectrum. The artificial smooth bump at the wavelength of $H{\alpha}$ in calibration vectors is introduced by the spline interpolation over the masked region. The emission line feature around 6708 $\rm \AA$ is due to that the model spectra underpredict the strength of LiI 6708 $\rm \AA$ absorption line.}
\label{}
\end{figure}

\begin{figure*}
\center
\includegraphics[width=\textwidth]{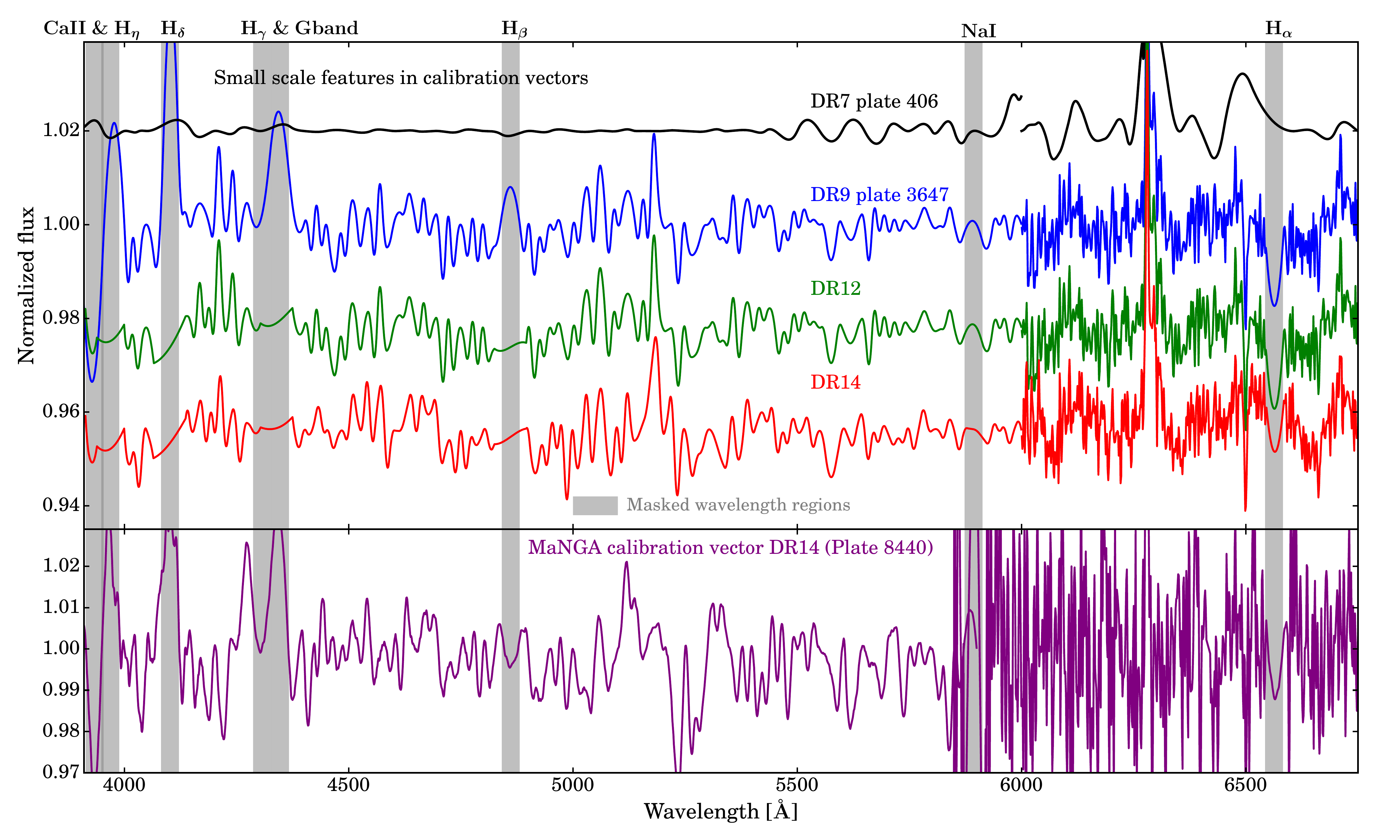}
\caption{Small scale features in calibration vectors. Upper panel: small scale features in a calibration vector of plate 406 from DR7 (black) and plate 3647 from DR9 (blue), DR12 (green), and DR14 (red) data release.
Lower panel: small scale features in a calibration vector of MaNGA survey (plate number 8440). The gray bands indicate the masked wavelength regions. Note that the $H{\alpha}$ absorption bump remains uncorrected.}
\label{fig:balmer_evolution}
\end{figure*}


\begin{figure*}
\includegraphics[width=\textwidth]{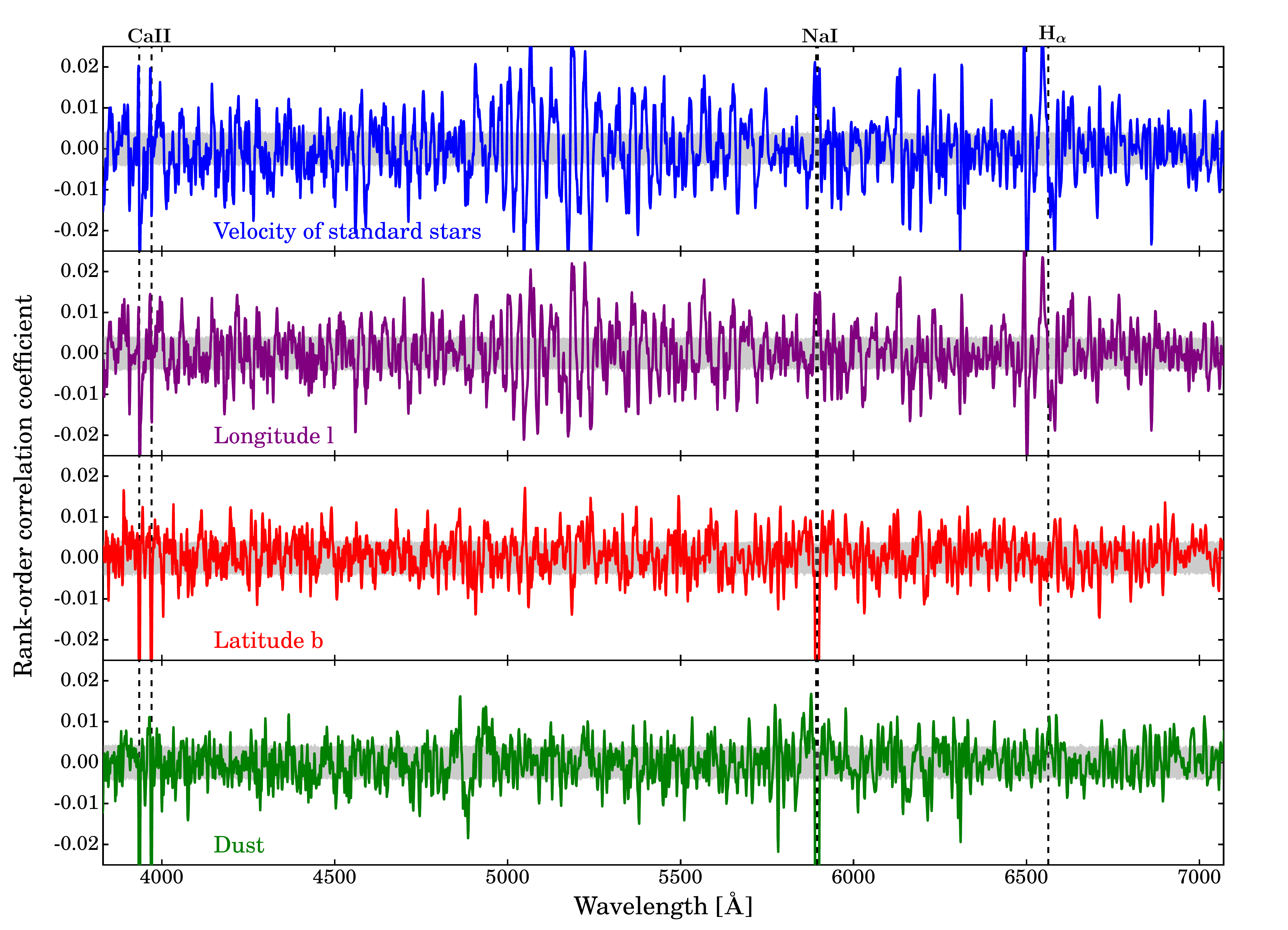}
\caption{Spearman's rank-order correlation spectra. The correlation spectra with velocity of the standard stars and longitude show significant oscillation features, which are originated from the velocity dependence of calibration vectors. The grey bands indicate  $1 \sigma$ regions.}
\label{fig:C}
\end{figure*}

\subsection{Artificial hydrogen Balmer emission \& absorption lines imprinted on SDSS spectra}

We now focus on the spectral regions corresponding to the hydrogen Balmer lines. Figure~\ref{fig:comparison} shows the behavior of the DR12 composite absorption spectrum significantly differs from that of the theoretical F-star spectra at those locations. This is expected as the strength of the stellar Balmer absorption line is orders of magnitude stronger than those of the metal lines. Given that these spectral regions are likely to result in substantial systematics in the calibration vector, no attempt is made to fit them accurately. Instead, the pipeline applies masks in the vicinity of the Balmer series and interpolates the estimated calibration vectors in those regions. When a spline interpolation is performed across a masked region, the final estimate will carry the signature of the local curvature of the original spectrum. At the location of $H\alpha$, the curvature is such that the interpolated calibration vector displays an absorption-like feature. This is illustrated in Figure~5 where we show 10 randomly selected calibration vectors in DR12. As can be seen, at the $H{\alpha}$ wavelength, there is a smooth artificial curving bump which appears in all the calibration vectors. For the majority of the objects, this artificial bump introduces an artificial absorption-like feature at the wavelength of $H{\alpha}$ which will be imprinted into the final science spectra. Its final amplitude will depend on the number of anchor points used in the spline interpolation of the calibration vector. This explains why the artificial $H{\alpha}$ feature is hardly noticeable in DR7 but is clearly present in DR9 and subsequent releases. 

This absorption-like feature was reported by \citet{ZhangHa} who analyzed the DR12 dataset. These authors interpreted it as $H{\alpha}$ absorption lines induced by gas in the halo of the Milky Way. \citet{Sethi2017} pointed out that such an amount of absorption is physically implausible. Our analysis now shows that this feature is artificial and fully explained by the limitations of the spectroscopic calibration procedure. As pointed out earlier, both the absorption features and the masks related to spectroscopic features of standard stars are expected to carry the signature of the star velocity distribution. Their position in the observer frame is expected to depend on the position in the sky and reflect the rotation of the Milky Way.

Over time, the data releases have treated the Balmer line regions in different ways. The presence of the artificial curvature was first reported by \citet{Busca2013} in the context of the DR9 data release. To minimize this effect, a different scheme was used in DR12 \citep[][see their Table 2]{Alam2015} by using a linear function (instead of an iterative b-spline procedure) to interpolate the flux over the masked regions. Surprisingly, we observe that this data reduction change was only applied to the Balmer $\beta$, $\gamma$ and $\delta$ lines but {\it not} applied to the Balmer $\alpha$ line. For the latter one, the original iterative spline interpolation scheme has been used all along. As a result, an absorption-like feature at the location of Balmer $\alpha$ is found in SDSS data releases 9 up to now, i.e., the latest data release 14 \citep{Abolfathi2017}. To illustrate this, we show examples of calibration vectors for SDSS BOSS DR9 \citep{Dawson2013, Ahn2012}, DR12 \citep{Alam2015}, eBOSS DR14 \citep{Dawson2016, Abolfathi2017} data release as well as calibration vectors for the MaNGA survey \citep{Bundy2015} DR14 data release \citep{Abolfathi2017} in Figure~\ref{fig:balmer_evolution}. In the upper panel, we show small scale features in calibration vectors from the same plate (number 3647) processed by the DR9 (blue), DR12 (green), and DR14 (red) pipelines. The spectra are obtained by normalizing the large scale features in calibration vectors with a cubic b-spline with break points separated by $50\,\rm \AA$. The three spectra are mostly identical while the smooth curving features at the wavelengths of Balmer series in DR9 disappear in DR12 and DR14. From DR12, the pipeline corrects the features with linear interpolation across wavelength regions with Balmer $\beta$, $\gamma$ and $\delta$ lines \citep{Alam2015}. One can also observe the change in the median residual spectra of DR9 and DR12 in Figure~1. However, the $H{\alpha}$ feature remains uncorrected from DR9 to DR14. 
We also show a DR7 calibration vector (black) in which most of the wiggles are absent. As pointed previously, this is due to the fact that the DR7 pipeline interpolates the calibration vectors using an effective scale larger than that used in subsequent data releases.

In the lower panel of the figure, we show an example of small scale features in DR14 MaNGA calibration vectors obtained with the same method. We note that the MaNGA survey uses a different pipeline for calibration. Therefore, the systematic features differ from the features in quasar spectra. However, we find that the imprints of Balmer series, including the $H{\alpha}$ feature, similar to the ones in DR9 BOSS calibration vectors. This illustrates that the current dataset of the MaNGA survey may suffer from similar systematic features as those found in BOSS/eBOSS quasar spectra. 

\subsection{Correlation spectra}
We can obtain further insight into the origin of the systematic spectroscopic fluctuations and their dependencies on observables by measuring cross-correlations with an external observable $X$. As we are looking for unknown systematics, we use a Spearman rank-order coefficient to be generic:
\begin{equation}
\label{eq:C}
C_{\delta_F\,X}=\left \langle {\rm Rank}(\delta_F)\,\cdot\,{\rm Rank}(X) \right \rangle(\lambda)\;.
\end{equation}
We have estimated the cross-correlation with observables $X$, such as air-mass, seeing, sky location, ISM column density, etc. The results for DR12 are shown in Figure~\ref{fig:C}. To enhance the signal to noise ratio, we smooth individual quasar spectra with a 1-pixel Gaussian kernel before calculating the correlation spectra. The corresponding uncertainties are obtained by bootstrapping the quasar sample 200 times.
As can be seen in the figure, correlations with Galactic longitude reveals many pixels with significant values. They are found around the $5000 \ \rm \AA$ and the $H{\alpha}$. The same spectral regions show even stronger correlation coefficients with the velocity of the standard stars used in each SDSS spectroscopic plate. This dependence reflects the connection between the systematic fluctuations in the composite absorption spectrum and the properties of the observed standard stars. We also detect correlations with Galactic latitude and dust column density at the location of the Calcium and Sodium absorption lines. This is expected as they all trace the column density of the ISM.

\section{Summary}

Large spectroscopic surveys provide us with samples of spectra, offering appealing opportunities for statistical analyses. They enable the detection of emission and absorption signals buried into measurement noise and/or population variance. Pushing the accuracy of statistical studies with such datasets requires understanding their limitations by systematic effects. We have studied these limitations in the context of absorption measurements with the Sloan Digital Sky Surveys I, II, III and IV (up to data release 14).

Using quasar spectra, we have shown
that changes in the data reduction strategy throughout different data releases have led to a better accuracy at long wavelengths but a degradation at short wavelengths with the emergence of systematic spectral features with an amplitude of about one percent. We have shown that these features originate from the spectroscopic calibration process which relies on the comparison between observed and modeled F-stars. In particular, we find that
\begin{itemize}
    \item The best-fit F-star models tend to over-predict the strength of metal lines and under-predict that of Lithium. Such effects can be seen at the plate-level, based on 20 reference stars.
    \item Different levels of residuals are found at the location of the hydrogen Balmer $\beta$, $\gamma$ and $\delta$ across the different data releases, due to changes in the type of interpolation scheme used to describe the corresponding masked regions. Surprisingly, the treatment of the Balmer $\alpha$ line has not changed up to the latest data release (DR14) and results in the presence of an artificial absorption line.

    We have shown that the H$\alpha$ absorption-like feature reported by \citet{ZhangHa} and interpreted as circum-galactic absorption is an artificial feature introduced by the calibration process.
    
    \item All these systematic features imprint a systematic pattern in all the reduced spectra. This pattern also reflects the velocity distribution of the standard stars across the sky and therefore carries their velocity distribution.
\end{itemize}
Our analysis shows that a more accurate calibration procedure can be implemented.
Having learned from the SDSS experience, our advice for the design and data reduction strategy of future surveys is as follows:
\begin{itemize}
    \item Composite continuum-normalized spectra for samples of sources should be systematically measured at different steps of the data reduction to check for the existence of systematic effects. In addition, cross-correlations between flux fluctuations and various observables such as dust reddening, Galactic latitude \& longitude, standard star velocity, etc. should be used as quality checks. Informative tests can be performed using only a limited set of reference sources (about 20 in the case of SDSS).

   \item Spectroscopic calibration procedures relying on reference stars should carefully investigate the required accuracy of the chosen library of stellar spectra (whether it is defined theoretically of empirically, e.g. \citet{Yan2017}), the sampling resolution of the stellar parameters and the 
   characterization of the PSF to ensure that the best fit spectra do not induce systematic features.
   
   \item A Sloan-like calibration process uses small-scale spectroscopic information to derive large-scale calibration. The corresponding calibration vector needs to be sampled on a carefully chosen wavelength scale such that small-scale features potentially left by imperfect modeling of the reference stellar spectra do not propagate to the science spectra.
   
    \item While SDSS selects F-stars as calibration standards, it is also possible to use other sources as calibration stars. For example, an intriguing set of quasi-perfect black-body objects has been reported by \citet{Suzuki2017} recently. These objects have no absorption features and their SEDs can be described with just two parameters. It will be interesting to characterize  their properties and estimate their number density on the sky. If they have a sufficient number density, they can be excellent candidates as calibration standards. 
\end{itemize}
Upcoming spectroscopic surveys such as SDSS-V, DESI, PFS,  
4MOST \citep{4most} and MOONs \citep{Moons} would benefit from such considerations.

\hskip10mm
\acknowledgements

We thank the referee, Michael Bessell, for his constructive report.
We thank Guangtun Zhu, who made his absorption spectra available to us in a digital form. We thank Masataka Fukugita, Renbin Yan, and Bruce~Draine for useful discussions. 
B.M. acknowledges support from NASA ADAP NNX16AF64G.
D.P. acknowledges support from ISF grant  541/17.

Kavli IPMU is supported by World Premier International Research Center Initiative of the Ministry of Education, Japan.

Funding for the SDSS and SDSS-II has been provided by the Alfred P. Sloan Foundation, the Participating Institutions, the National Science Foundation, the U.S. Department of Energy, the National Aeronautics and Space Administration, the Japanese Monbukagakusho, the Max Planck Society, and the Higher Education Funding Council for England. The SDSS Web Site is http://www.sdss.org/. 

Funding for SDSS-III has been provided by the Alfred P. Sloan Foundation, the Participating Institutions, the National Science Foundation, and the U.S. Department of Energy Office of Science. The SDSS-III web site is http://www.sdss3.org/. SDSS-III is managed by the Astrophysical Research Consortium for the Participating Institutions of the SDSS-III Collaboration including the University of Arizona, the Brazilian Participation Group, Brookhaven National Laboratory, Carnegie Mellon University, University of Florida, the French Participation Group, the German Participation Group, Harvard University, the Instituto de Astrofisica de Canarias, the Michigan State/Notre Dame/JINA Participation Group, Johns Hopkins University, Lawrence Berkeley National Laboratory, Max Planck Institute for Astrophysics, Max Planck Institute for Extraterrestrial Physics, New Mexico State University, New York University, Ohio State University, Pennsylvania State University, University of Portsmouth, Princeton University, the Spanish Participation Group, University of Tokyo, University of Utah, Vanderbilt University, University of Virginia, University of Washington, and Yale University.

Funding for the Sloan Digital Sky Survey IV has been provided by the Alfred P. Sloan Foundation, the U.S. Department of Energy Office of Science, and the Participating Institutions. SDSS acknowledges support and resources from the Center for High-Performance Computing at the University of Utah. The SDSS web site is www.sdss.org. SDSS-IV is managed by the Astrophysical Research Consortium for the 
Participating Institutions of the SDSS Collaboration including the 
Brazilian Participation Group, the Carnegie Institution for Science, 
Carnegie Mellon University, the Chilean Participation Group, the French Participation Group, Harvard-Smithsonian Center for Astrophysics, 
Instituto de Astrof\'isica de Canarias, The Johns Hopkins University, 
Kavli Institute for the Physics and Mathematics of the Universe (IPMU) / 
University of Tokyo, Lawrence Berkeley National Laboratory, 
Leibniz Institut f\"ur Astrophysik Potsdam (AIP),  
Max-Planck-Institut f\"ur Astronomie (MPIA Heidelberg), 
Max-Planck-Institut f\"ur Astrophysik (MPA Garching), 
Max-Planck-Institut f\"ur Extraterrestrische Physik (MPE), 
National Astronomical Observatories of China, New Mexico State University, 
New York University, University of Notre Dame, 
Observat\'ario Nacional / MCTI, The Ohio State University, 
Pennsylvania State University, Shanghai Astronomical Observatory, 
United Kingdom Participation Group,
Universidad Nacional Aut\'onoma de M\'exico, University of Arizona, 
University of Colorado Boulder, University of Oxford, University of Portsmouth, 
University of Utah, University of Virginia, University of Washington, University of Wisconsin, 
Vanderbilt University, and Yale University.

\end{document}